# PHYSICAL INTEGRITY ATTACK DETECTION OF SURVEILLANCE CAMERA WITH DEEP LEARNING BASED VIDEO FRAME INTERPOLATION

JONATHAN PAN

Nanyang Technological University, Singapore
E-MAIL: JonathanPan@ntu.edu.sg

**Abstract:**

Surveillance cameras, which is a form of Cyber Physical System, are deployed extensively to provide visual surveillance monitoring of activities of interest or anomalies. However, these cameras are at risks of physical security attacks against their physical attributes or configuration like tampering of their recording coverage, camera positions or recording configurations like focus and zoom factors. Such adversarial alteration of physical configuration could also be invoked through cyber security attacks against the camera's software vulnerabilities to administratively change the camera's physical configuration settings. When such Cyber Physical attacks occur, they affect the integrity of the targeted cameras that would in turn render these cameras ineffective in fulfilling the intended security functions. There is a significant measure of research work in detection mechanisms of cyber-attacks against these Cyber Physical devices, however it is understudied area with such mechanisms against integrity attacks on physical configuration. This research proposes the use of the novel use of deep learning algorithms to detect such physical attacks originating from cyber or physical spaces. Additionally, we proposed the novel use of deep learning-based video frame interpolation for such detection that has comparatively better performance to other anomaly detectors in spatiotemporal environments.

## 1. Introduction

Surveillance cameras are an integral part of the smart infrastructure to provide security video surveillance to monitor and record areas of interest. Surveillance cameras are typically deployed in public areas like shopping malls, roadways, transportation hubs and also transportation platforms like trains. These surveillance cameras are operated by their human operators, processed by video analytics tools to detect activity or object of interest and videos are stored into video storage medium for later retrieval for post incident analysis like incident investigation or digital forensics. For such video surveillance to support their intended usages, the integrity of physical configuration of these surveillance cameras needs to be upkept and prevent any unauthorized changes. Configurations like the camera's position or placement, Pan-Tilt-Zoom (PTZ) configuration and their ability to maintain the line-of-sight of the area of surveillance.

There are however threats and vulnerabilities to these surveillance camera's physical configuration. Such malicious attack may occur in cyber or physical realms. Cyber induced form of attacks would first exploit the software vulnerability of the targeted camera and consequentially cause adversarial digitally-enabled alteration of the physical configuration. Physical tampering of cameras could be the simpler or less technically complex forms of attack that only requires the attacker to have physical reach to the camera to effect change to the configuration [17]. The effects of such physical tampering could result in the disruption of video stream or corrupt surveillance coverage. Conventional forms of detection mechanism of physical alteration would require human operators to manually study or observe video feeds. When there are large numbers of surveillance cameras involved, such monitoring efforts are inefficient and ineffective. Site inspections of camera deployments would be labour intensive and have long lead-time to detect such occurrences.

This research proposes two novelties. The first is the application of Artificial Intelligence based Deep Learning algorithms to detect anomalous tampering of the physical configuration of surveillance cameras. The second is the proposition of a novel approach to video anomaly detection using video frame interpolation applied to spatiotemporal video feeds. Unlike most video anomaly detector, this is not based on visual optics flow. With this technique, it would provide an out-of-band visual layer approach to detect physical tampering of surveillance camera as a human operator could on video stream that is susceptible to varying degree of environmental changes within the surveillance area.

The next section of this paper provides background information about the forms of Cyber Physical attacks against surveillance and video frame interpolation. This is followed by related research work in detecting such form of attack and the popular types of deep learning video anomaly detectors. The description of the model is covered, followed by description of experiments and analysis done. The paper concludes with a conclusion and discussion about future research directions.

## 2. Background Information

### 2.1. Cyber Physical Attacks Against Surveillance Cameras

The physical dimension of surveillance cameras is a prime candidate for an attack vector. Such physical attacks would include tampering of the cameras' physical configuration like changing of the cameras' position, adjusting the PTZ, blocking the cameras' line of sight or fiddling with the cameras' lens focal or zoom settings. These cameras could easily be physically damaged that would totally disrupt the video feeds. They could also be at risks of environmental based disruption attack to the dependent power supply, communication links or physical mountings used with these cameras [11]. While there are physical tamper protection solutions available, there are limits to such physics-based protection and less so for detection capabilities. However, such tampering could easily be detected by an observant human operator.

Surveillance cameras, like all Internet of Things (IoT) devices, are also at risk to a wide range of cyber threats. As these IoT devices connect digitally to provide video feeds and receive administrative control commands, they are also vulnerable to cyber-attacks that could have physical consequences. Unauthorized privileged remote access to these surveillance cameras could alter the configuration of the vulnerable cameras preventing them from performing their intended surveillance functions or coverage [10]. Such forms of Cyber Physical attacks [15] originating primarily in cyber dimensions with physical consequence is likened to the Stuxnet malware attack that altered the physical operations of the targeted uranium enrichment facility.

With the varied forms of cyber, physical or combined forms of attacks that could affect the configuration integrity of these IoT cameras, there are limits to detect such occurrence to prevent or recover from such attacks. The cyber detection mechanisms like Intrusion Detection System (IDS) would be effective against cyber-attacks; however, they are less capable of detecting physical attacks. Conversely physical attacks could cause significant cyber consequential impact to the intended digital operations of these IoT cameras. An approach to detect cyber physical attack variants would be to have an operator manually study the video feeds and identify any observable unintended changes to the video feeds. Hence this preliminary research work attempts to automate human's abilities to analyze video feeds for any observable unintended changes induced by physical alterations while ignoring normal contextual environmental changes.

### 2.2. Video Frame Interpolation

Interpolation, within the mathematical field of numerical analysis, involves the approximated proposition of missing data points from a range of known data points. It has been used generate high resolution images from low resolution images [12]. Video frame interpolation involves the synthesis of nonexistent frames in-between the frames. It is typically used in rendering, compression and increasing video quality as part of computer vision and digital image processing. A conventional approach for frame generation is to use motion estimation-based methods that entails the computation of missing frames through the estimation of optical flows [14] or moving gradient techniques [13]. There are significant interests in applying deep convolutional neural networks to video frame interpolation. Mathieu et al. [7] used Mean Squared Error (MSE) as its loss function and multi-scale architecture to sharpen video frames. Liu et al [21] adopted optical flow-based networks to synthesis new video frames. However, there is none applied in the context of anomaly detection. Hence our research novelty is the application of Deep Learning based Video Frame Interpolation as an anomaly detector.

## 3. Related Work

### 3.1. Physical Integrity Tampering Detection

Giraldo et al. [1] surveyed the research trends in developing detection solutions to detect varied forms of 'physics-based' attacks against Cyber Physical Systems. They observed many forms of attacks against the 'Trust' factor or the integrity of physical components of Cyber Physical Systems would likely have observable attack effects. However, the problem is that such attack induced deviation effects may not be monitored hence detection alarms would not be raised.

Valente and Cardenas [4] proposed the use visual challenges like QR codes to verify the freshness of camera video footage. The intent is to detect forms of adversarial video attacks that included moving cameras to point to different area. However, this detection solution would detect anomalies as long as the visual challenge remain visible to the camera. Zhang et al. [3] also observed that the physical environmental conditions are typically unique to the IoT's deployment hence they are diverse. Hence any detection solution would need to be contextually tuned to the environment to minimize false alarms yet optimally sensitive to detect environmental integrity attacks. They also concluded that the study of dealing with threats originating from physical environment is understudied. This research work addresses the need to have a contextually tuned detector and contributed to this understudied area.

3.2. Deep Learning based Anomaly Detectors

Deep Learning algorithms have been very popular lately in their application to anomaly or fault detection. This is due to their ability to learn features (including video feeds) from raw data with their deep architectures compromising of layers of non-linear data processing units called neural networks. There are a variety of Deep Learning based methods used for such detection [9]. According to Ravi et al., they are broadly classed into the following three.

- **Autoencoder for reconstruction** that uses representational learning approach that models normal behavior of surveillance videos through linear and non-linear transformations of image flows from video footages.
- **Predictive modeling** like Convolutional Recurrent Neural Network models attempts to predict future video frames from current temporal spatial video frames based on conditional distribution $P(x_t |(x_{t-1}, x_{t-2}, \ldots x_{t-p}))$.
- **Generative models** like Variational Autoencoders (VAE) attempts to model the likelihood of normal video feeds through the inclusion of stochastic variations.

Anomaly detection with these unsupervised algorithms are done through the measure of variation of the reconstructed and actual video frame pairs against defined threshold.

4. Models

4.1. Base Deep Learning Algorithm

In our research, we applied the Convolutional Long-Short Term Memory (ConvLSTM) [16] as the base architecture for our anomaly detector for physical attacks against the Cyber Physical surveillance camera. The ConvLSTM extends the fully connected Long-Short Term Memory (LSTM) to include convolutional structures in both the input-to-state and state-to-state transitions. With convolutional kernels, the ConvLSTM ingests the video frame images with higher dimensions of spatial temporal associations as inputs. The following are the mathematical expressions for the ConvLSTM.

$$I_t = \sigma(W_{xi} * X_t + W_{hi} * H_{t-1} + W_{ci} \circ C_{t-1} + b_i) \quad (1)$$
$$F_t = \sigma(W_{xf} * X_t + W_{hf} * H_{t-1} + W_{cf} \circ C_{t-1} + b_f) \quad (2)$$
$$C_t = f_t \circ C_{t-1} + i_i \circ \tanh(W_{xc} + X_t + W_{hc} * H_{t-1} + b_c) \quad (3)$$
$$O_t = \sigma(W_{xo} * X_t + W_{ho} * H_{t-1} + W_{co} \circ C_{t-1} + b_o) \quad (4)$$
$$H_t = o_t \circ \tanh(V_t) \quad (5)$$

In comparison to LSTM, the ConvLSTM has '$*$' operators that performs convolution operation while '$\circ$' is the Hadamard product for the element-wise product of matrices. The rest of network structure performs, like the LSTM network, processes the input, forget, cell, output and hidden state computations for each timestep denoted by $I, F, C, O$ and $H$ respectively with activation by $\sigma$ and convolutional filter computation with the sets of weights, $W$. Each state is represented as a matrix representing a video frame. ConvLSTM has been used extensively as the base architecture for video anomaly detection.

4.2. Anomaly Detector Algorithms

With the base ConvLSTM, we applied the popular three forms of anomaly detectors [9] namely Autoencoder [18], Predictive [19] and VAE [20] used as comparative models to our model.

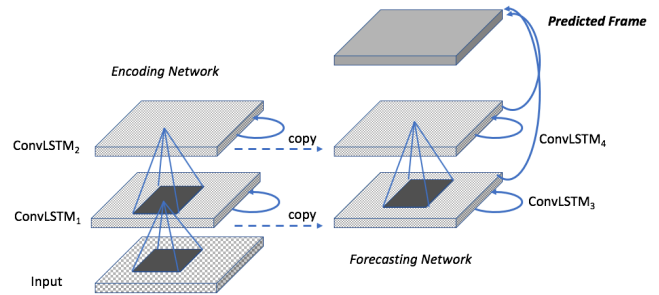

**FIGURE 1.** ConvLSTM Predictor

Our model is similarly structured to the Predictive ConvLSTM [9] shown in Figure 1. The Predictive model takes in a sequence of video frames $(x_{t-J}, x_{t-J-1}, \ldots, x_{t-1}, x_t)$, where $J$ is the length of the LSTM's timestep, performs unsupervised feature extraction through an encoder which then feeds to a decoder that is trained to deconvolute and predict the next video frame $(x_{t+1})$ through reconstructive approximation as illustrated mathematically below.

$$\begin{aligned}\hat{x}_{t+1} &= p(x_{t+1}|x_{t-J}, x_{t-J-1}, \ldots x_{t-1}, x_t) \\ &\approx p(x_{t+1}|f_{encoding}(x_{t-J}, x_{t-J-1}, \ldots x_{t-1}, x_t)) \\ &\approx g_{interpolate}(f_{encoding}(x_{t-J}, x_{t-J-1}, \ldots x_{t-1}, x_t))\end{aligned} \quad (6)$$

Our model, based on deep learning convolutional interpolation [8], involves approximating the interpolation of the missing video frame $\widehat{\theta_{t-1}}(x,y)$ from the pair of adjacent video frames available $\theta_{t-2}(x,y)$ and $\theta_t(x,y)$.

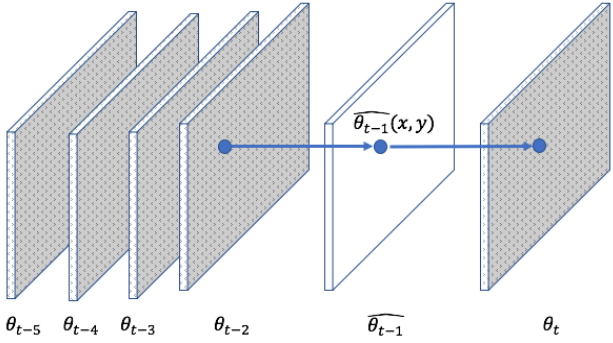

**FIGURE 2.** Video Frame Interpolation

The encoding ConvLSTM network ingests the input sequence of video frames $(x_{t-J}, x_{t-J+1}, \ldots, x_{t-2}, x_t)$ into its hidden layers of ConvLSTM with one video frame excluded ($x_{t-1}$). The Interpolation ConvLSTM network will unfold the hidden state to reconstruct the missing video frame through interpolation.

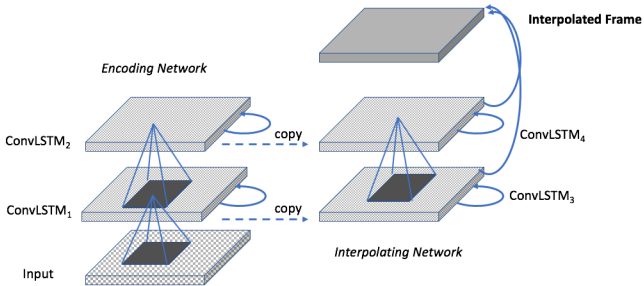

**FIGURE 3.** ConvLSTM Interpolator

$$\begin{aligned}\hat{x}_{t-1} &= p(x_{t-1}|x_{t-J}, x_{t-J-1}, \ldots x_{t-2}, x_t) \\ &\approx p(x_{t-1}|f_{encoding}(x_{t-J}, x_{t-J-1}, \ldots x_{t-2}, x_t)) \\ &\approx g_{interpolate}(f_{encoding}(x_{t-J}, x_{t-J-1}, \ldots x_{t-2}, x_t))\end{aligned} \quad (7)$$

We adopted the same Mean Square Error (MSE) loss function used by Mathieu et al. [7], in applying deep learning on video interpolation to solve blurriness in frame prediction, to train the interpolation model to reconstruct the missing frame. During inference, the model computes the interpolated missing video frame against the actual video frame that it is given using MSE where $\hat{\theta}$ is the inferred interpolated output, $\theta$ is the actual value and $p$ is the total number of pixels contained in that frame.

$$e = \frac{1}{p}\sum_{i=1}^{p}(\hat{\theta}_{ki} - \theta_{ki})^2 \quad (8)$$

A high MSE that exceeds a defined threshold between the two frames (computed interpolated reconstruct and the actual) indicates the notable change noted by the model hence indicating an anomalous occurrence.

## 5. Methodology and Analysis

### 5.1. Dataset

For our video training data, we used two sets of video surveillance camera footages. Both sets are stationary train surveillance cameras with people moving through the train carriages with segments of video footage with passengers performing some atypical activities like hugging and fighting in a moving train with constantly changing environmental background observed through the window screens of the train carriages. Both surveillance camera datasets were captured as part of Eureka Celtic Initiative supported by four European countries [5] focused on improving train passenger security video analytics. Two cameras datasets are cameras from different locations as shown below.

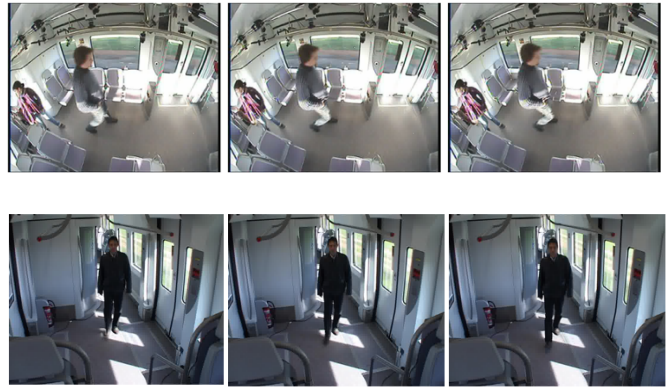

**FIGURE 4.** Surveillance Cameras on Moving Train

We trained all models to tune them to learn the spatiotemporal features of both environments. We also created a sets of test videos from both cameras that simulated physical attacks. The training dataset accounted about 90% of the video footage and 10% for testing dataset. The simulated attacks took the following forms of attacks with ten randomly selected time instances for each attack hence having imbalance class distribution of normal video footages and video segments with attack effects.

- **Blocking camera's view** that simulated by masking out the video footage with a black image. This attack mimics a physical attack of covering the camera or perhaps physically making the camera's capture lens opaque through application of water based paint.

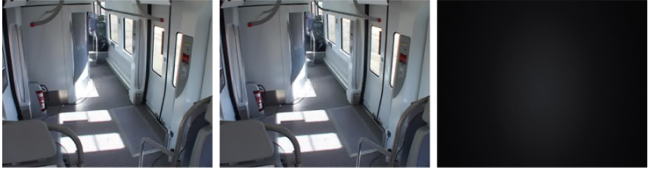

**FIGURE 5.** Rightmost video frame simulates blocked view

- **Adjusting the zoom configuration of the camera** that was simulated by cropping specific video frames. This simulation simulates the attack of physically adjusting the zoom factor of the camera configuration hence causing the camera to limit its viewing angle and obscuring parts of the intended surveillance coverage.

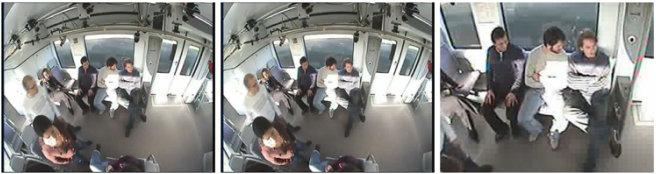

**FIGURE 6.** Rightmost video frame simulates zoom-in

- **Altering the focus of the camera** that was simulated by applying image filters (blurring) to specific video frames. The simulation seeks to mimic the physical manual adjustments of focal lens of the camera hence affecting the clarity of the captured video feed.

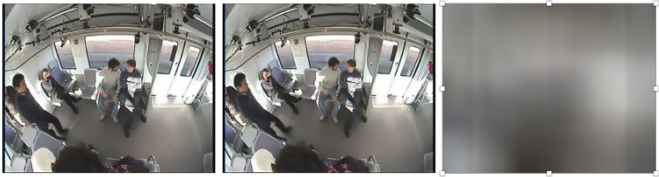

**FIGURE 7.** Rightmost video frame simulates blurring effect

- **Adjusting the camera's viewing angle** that was simulated by shifting the viewing area of the recorded video footage. This attack simulates a physical left tilt of the camera position. The effects of the change of viewing area could obscure area of the view that may be important to closely monitor.

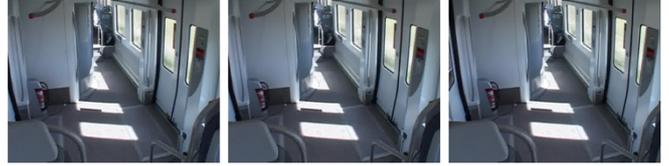

**FIGURE 8.** Rightmost video frame simulates small left-shift

### 5.2. Results and Analysis

All models were trained and tested with the four sets of test scenarios. An evaluation criterion based on True Positive, True Negative, False Positive, False Negatives were collected. Additionally, the F1, Precision and Recall values were computed. Hence a total of 32 test evaluation results were collected with the four test scenarios with two sets of previously unseen camera videos across four models including our interpolation anomaly detector model. The following are the results summarized and organized by the four test scenarios.

**TABLE 1.** Blocking Camera's View

| Model | Evaluation | | | |
|---|---|---|---|---|
| | *TP/TN/FP/FN* | *F1* | *Precision* | *Recall* |
| Predictor | 20/7828/5/0 | 0.89 | 0.80 | 1.00 |
| **Interpolator** | **20/7832/1/0** | **0.98** | **0.95** | **1.00** |
| VAE | 20/7822/11/0 | 0.78 | 0.65 | 1.00 |
| AE | 0/7833/0/20 | N/A | N/A | 0.00 |

**TABLE 2.** Adjust Camera's Zoom

| Model | Evaluation | | | |
|---|---|---|---|---|
| | *TP/TN/FP/FN* | *F1* | *Precision* | *Recall* |
| Predictor | 20/7828/5/0 | 0.89 | 0.80 | 1.00 |
| **Interpolator** | **20/7832/1/0** | **0.98** | **0.95** | **1.00** |
| VAE | 20/7821/12/0 | 0.77 | 0.63 | 1.00 |
| AE | 4/7833/0/16 | 0.33 | 1.00 | 0.20 |

**TABLE 3.** Alter Camera's Focus

| Model | Evaluation | | | |
|---|---|---|---|---|
| | *TP/TN/FP/FN* | *F1* | *Precision* | *Recall* |
| **Predictor** | **18/7828/5/2** | **0.84** | **0.78** | **0.90** |
| Interpolator | 10/7826/7/10 | 0.54 | 0.59 | 0.50 |
| VAE | 3/7821/12/17 | 0.17 | 0.20 | 0.15 |

| Model | Evaluation | | | |
|---|---|---|---|---|
| | TP/TN/FP/FN | F1 | Precision | Recall |
| AE | 0/7833/0/20 | N/A | N/A | 0.00 |

**TABLE 4.** Shift Camera – Camera 1

| Model | Evaluation | | | |
|---|---|---|---|---|
| | TP/TN/FP/FN | F1 | Precision | Recall |
| Predictor | 19/7826/7/1 | 0.83 | 0.73 | 0.95 |
| **Interpolator** | **19/7831/2/1** | **0.93** | **0.90** | **0.95** |
| VAE | 10/7822/11/10 | 0.49 | 0.48 | 0.50 |
| AE | 0/7831/2/20 | N/A | 0.00 | 0.00 |

The test results indicate that the anomaly detector with video frame interpolation performed the best compared to the other models for physical attacks for three out of the four test scenarios. It was second best for the simulated tests of adjusting the camera's focus.

As part of testing, we rendered the interpolated video frame of our our ConvLSTM Interpolator model to study the reconstructed video frame. We compared the interpolated video frames when there were nobody present in the train carriage and where there was.

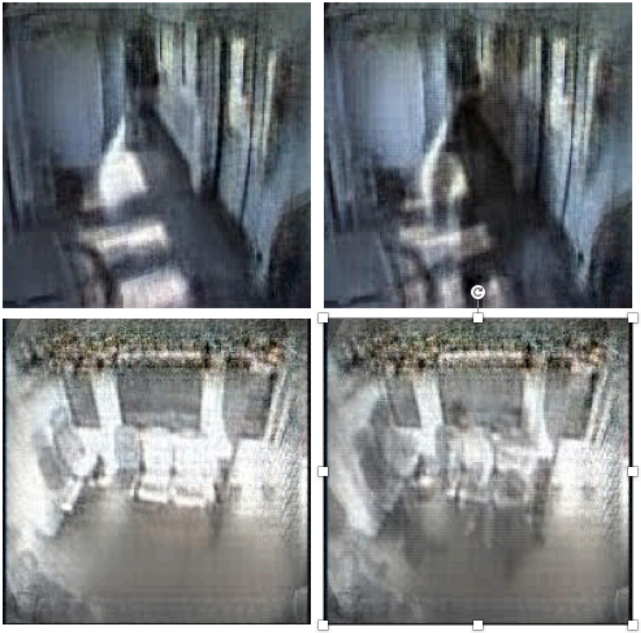

**FIGURE 9.** Reconstructed video frame by our ConvLSTM Interpolator when the train carriage is empty (left) and has passengers (right)

Notably the left images of the reconstructed video frames has a definitive image representation of the environmental settings of the train carriage. Where there are image segments in the video frame that may contain transient objects like moving or seated passengers present in the surveillance footage, the model would generate a silhouette form of those objects. It is apparent that our model ConvLSTM Interpolator has tuned itself to environmental settings of the surveillance coverage.

## 6. Conclusion and Future Directions

Surveillance cameras are at risks of physical integrity attacks and it is an understudied area. There is serious functional surveillance impact from attacks on the physical configurations of these surveillance cameras that could originate from cyber or physical spaces. When such attack occurs and remains undetected, the video recordings of the attacked cameras could be rendered useless when they are analyzed for investigative purposes or digital video forensics.

This research work demonstrates the application of deep learning algorithms to perform visual layer inspection or monitoring of surveillance video feeds to spot any observable forms of surveillance coverage integrity attacks. Additionally, this work further extends the typical classes of deep learning based anomaly detection algorithms to include a novel form using video frame interpolation.

Future research work will involve applying the detection approach to actual surveillance camera feeds to detect real attacks. Also, to apply it on archived video recordings to detect any past physical tampering or attack occurrences. Finally, to extend the application of interpolation based anomaly detection to other domains beyond visual anomalies.